\documentclass[superscriptaddress,nofootinbib,amsmath,amssymb,
 aps,prd,onecolumn]{revtex4-2}

\usepackage{graphicx,subfigure,ulem}
\usepackage{dcolumn}
\usepackage{bm}
\usepackage{natbib} 
\usepackage[colorlinks=true,citecolor=blue,linkcolor=blue,urlcolor=blue]{hyperref}
\setcitestyle{square, numbers, comma, sort&compress} 
\usepackage{multirow}

\begin{document}

\title{Susceptibilities and Taylor coefficients of magnetic QCD from  perturbation theory}

\author{Eduardo S. Fraga}
\email{fraga@if.ufrj.br}
\affiliation{Instituto de F\'isica, Universidade Federal do Rio de Janeiro, Caixa Postal 68528, 21941-972, Rio de Janeiro, RJ, Brazil}

\author{Let\'icia F. Palhares}%
 \email{leticia.palhares@uerj.br}
 \affiliation{Universidade do Estado do Rio de Janeiro, Instituto de F\'isica, Departamento de F\'isica Te\'orica, Rua S\~ao Francisco Xavier 524, 20550-013 Maracan\~a, Rio de Janeiro, Brasil}
 
\author{Tulio E. Restrepo}
 \email{trestre2@central.uh.edu}
\affiliation{Department of Physics, University of Houston, Houston, TX 77204, USA}

\affiliation{Instituto de F\'isica, Universidade Federal do Rio de Janeiro, Caixa Postal 68528, 21941-972, Rio de Janeiro, RJ, Brazil}

\begin{abstract}
We compute the coefficients $c_2(T,B)$ and $c_4(T,B)$ of the Taylor expansion for the pressure in powers of $\mu_B/T$ in the presence of a large magnetic field within perturbative QCD at finite temperature and baryon density up to two-loops for $N_f=3$ flavors with physical quark masses.  We also present results for the excess of pressure, baryon density and baryon number susceptibility as functions of $\mu_B$, as well as susceptibilities as functions of the temperature in the $\{ \mu_B,\mu_Q,\mu_S \}$ basis. Our results can be directly compared to recent lattice QCD data.
Even though current lattice results do not overlap with its region of validity, perturbative results seem to be compatible with those obtained on the lattice for large temperatures.
\end{abstract}

\maketitle

\section{Introduction}
\label{introduction}

The phase diagram of strong interactions is ultimately to be constructed from in-medium quantum chromodynamics (QCD), the fundamental theory for hadronic matter. Besides temperature and different chemical potentials, the inclusion of an external magnetic field as one of the control parameters has proven to be phenomenologically relevant in different scenarios \cite{Adhikari:2024bfa}, from the astrophysics of compact stars \cite{Duncan:1992hi,Thompson:1993hn,Kouveliotou:1998ze} and binary neutron star mergers \cite{Jiang:2025ijp,Most:2025kqf} to non-central, high-energy heavy ion collisions \cite{Kharzeev:2007jp,Skokov:2009qp,Voronyuk:2011jd,Bzdak:2011yy,Deng:2012pc,Inghirami:2016iru,Roy:2017yvg}, as well as in the early universe \cite{Vachaspati:1991nm,Enqvist:1993np,Grasso:2000wj}.

First-principle calculations within magnetic QCD can be performed using Lattice QCD simulations \cite{DElia:2010abb,Bali:2011qj,Ilgenfritz:2012fw,Bali:2012zg,Bornyakov:2013eya,Bali:2013esa,Bruckmann:2013oba,Bali:2014kia,Endrodi:2015oba,DElia:2018xwo,DElia:2021tfb,DElia:2021yvk} (see Ref. \cite{Endrodi:2024cqn} for a recent review), perturbation theory \cite{Blaizot:2012sd,Ayala:2014uua,Ayala:2015bgv,Fraga:2023cef,Fraga:2023lzn,Fraga:2024klm}, hard thermal loop perturbation theory \cite{Rath:2017fdv,Haque:2017nxq,Karmakar:2019tdp,Bandyopadhyay:2017cle,Karmakar:2020mnj}, and in certain limits of QCD such as for a large number of colors $N_c$ \cite{Fraga:2012ev} and within chiral perturbation theory \cite{Colucci:2013zoa,Hofmann:2020dvz,Hofmann:2020ism,Adhikari:2021bou}. Of course, one can also adopt holographic models \cite{Preis:2010cq,Preis:2011sp,Preis:2012fh,Finazzo:2016mhm,Critelli:2016cvq} and a variety of effective models \cite{Fraga:2012rr,Kharzeev:2013jha,Andersen:2014xxa,Miransky:2015ava,Andersen:2021lnk,Adhikari:2024bfa}.

Recently, the QCD equation of state in the presence of an external magnetic field $B$ and nonzero baryon density $\mu_B$, besides the temperature $T$, was investigated via lattice simulations with $2+1$ dynamical staggered quarks at their physical masses \cite{Astrakhantsev:2024mat}. The simulations were performed using the method of the imaginary baryon chemical potential to circumvent the Sign Problem \cite{Aarts:2015tyj}. Their results indicate considerable enhancement of the coefficients for the expansion of the pressure in powers of $\mu_B/T$ by the presence of the magnetic field, considering values up to $eB=1.2$ GeV$^2$. The presence of the magnetic field also shifts their characteristic temperature dependence to lower values, which is consistent with the reduction of the critical temperature for the chiral transition and a strengthening of the transition as the magnetic field is increased \cite{Endrodi:2024cqn}.

In this paper we compute the coefficients $c_2(T,B)$ and $c_4(T,B)$ for the expansion of the pressure in powers of $\mu_B/T$ in the presence of a large magnetic field from first principles within perturbative QCD (pQCD) at finite temperature and baryon density up to two-loops for $N_f=3$ flavors with physical quark masses. Since we use perturbative QCD within the lowest-Landau level (LLL) approximation, the region of validity for our framework is given by $m_s \ll T \ll \sqrt{eB}$, where $m_s$ is the strange quark mass, $e$ is the fundamental electric charge, and $B$ is the magnetic field strength. We include the effects of the renormalization scale in the running coupling, $\alpha_s (T,\sqrt{eB})$, and running quark masses, since these effects have proven to be relevant for the resulting thermodynamics \cite{Freedman:1977gz,Farhi:1984qu,Fraga:2004gz,Kurkela:2009gj,Gorda:2021gha}. 

We also present results for the excess of pressure, baryon density and baryon number susceptibility as functions of $\mu_B$, as well as susceptibilities as functions of the temperature in the $\{ \mu_B,\mu_Q,\mu_S \}$ basis. Here, these are the chemical potentials for baryon number, charge and strangeness, respectively. These results, obtained from pQCD, can be directly compared to those obtained within Lattice QCD, keeping in mind that the ranges of temperature and magnetic field strength in current lattice simulations are still below the region of validity of the perturbative calculations in the LLL approximation. Nevertheless, the agreement seems to be very satisfactory. 
Even though one cannot address within the pQCD framework the region that lies close to the chiral transition, where the coefficients $c_i(T,B)$ exhibit a much richer structure, our perturbative results seem to be compatible with those obtained on the lattice for large temperatures and a magnetic field of $eB=1.2$ GeV$^2$. Finally, we provide predictions for the behavior of the coefficient $c_2(T,B)$ as a function of $eB$ for fixed (high) temperatures, which could be compared to the lattice once higher magnetic fields for this physical setup are attained in the future. The coefficient $c_4(T,B)$, on the other hand, is vanishingly small for large magnetic fields.

This work is organized as follows. In Section \ref{framework} we present the perturbative framework and a few details on the calculation of the pressure to two-loops, including the running of the coupling and strange quark mass, as well as the expressions for the coefficients $c_2(T,B)$ and $c_4(T,B)$. In Section \ref{results} we discuss our results for the coefficients $c_2(T,B)$ and $c_4(T,B)$, as well as the excess of pressure, baryon density and baryon number susceptibility as functions of $\mu_B$, and  susceptibilities as functions of the temperature in the $\{ \mu_B,\mu_Q,\mu_S \}$ basis, comparing them to the ones obtained using Lattice QCD simulations. Section \ref{summary} contains our summary and outlook.

\section{Theoretical framework}
\label{framework}

Let us start with the pressure of thermal QCD in the presence of high magnetic fields, which has been computed to two-loop order \cite{Blaizot:2012sd,Fraga:2023cef,Fraga:2023lzn}. The one-loop (free), contribution from the quark sector is given by the following renormalized expression (subtracting the pure vacuum term) \cite{Fraga:2012rr,Kharzeev:2013jha,Andersen:2014xxa,Miransky:2015ava}:

\begin{widetext}
\begin{align}
\begin{split}
 \frac{P_{\rm free}^q}{N_c}=&\sum_f\frac{(q_f B)^2}{2\pi^2}\left[\zeta^\prime(-1,x_f)-\zeta^\prime(-1,0)+\frac{1}{2}\left(x_f-x_f^2\right)\ln x_f+\frac{x_f^2}{2}\right]\\
 &+T\sum_{n,f}\frac{q_f B}{\pi}\left(1-\delta_{n,0}/2\right)\int \frac{dp_z}{2\pi}\bigg \{\ln\left(1+e^{-\beta\left[E(n,p_z)-\mu_f\right]}\right)+\ln\left(1+e^{-\beta\left[E(n,p_z)+\mu_f\right]}\right)\bigg \} \, ,
 \end{split}\label{P0}
 \end{align}
 \end{widetext}
where $E^2(n,p_z)=p_z^2+m_f^2+2q_f B n$, $x_f\equiv m_f^2/2q_f B$, $T=1/\beta$
is the temperature, $\mu_f$ is the quark chemical potential, $N_c$ is the number
of colors, $f$ labels quark flavors, $q_f$ is the quark electric charge, and
$n=0, 1, 2, \cdots$ stands for the Landau levels. In this expression, Matsubara
sums have already been performed in the medium
contribution\footnote{ One should notice that there is an inherent
arbitrariness in the renormalization procedure in the presence of a magnetic
background (see Refs.
\cite{Fraga:2008qn,Mizher:2010zb,Fraga:2012fs,Endrodi:2013cs,Haber:2014ula,
Avancini:2019wed,Avancini:2020xqe,Tavares:2021fik,Farias:2021fci} for a
discussion). In Eq. (\ref{P0}), all mass-independent terms were neglected and
the pure magnetic term goes to zero in the limit $m\to 0$. There are
renormalization procedures where other terms survive and the pure magnetic
expression diverges as $m\to 0$. This discrepancy in the renormalized expression
leads to differences in some physical quantities, e.g. the magnetization
\cite{Endrodi:2013cs}. It is important to note however, that the
susceptibilities and the expansion coefficients calculated below are independent
of the renormalization procedure, since the pure magnetic terms do not
contribute.}.

Taking the limit of very high magnetic fields ($m_s \ll T \ll \sqrt{eB}$), one ends up with the lowest Landau level (LLL) expression
\begin{widetext}
\begin{align}
\begin{split}
 \frac{P_{\rm free}^{\rm LLL}}{N_c}=&
 -\sum_f\frac{(q_fB)^2}{2\pi^2}\left[x_f\ln\sqrt{x_f}\right]+T\sum_{f}\frac{q_fB}{2\pi}\int \frac{dp_z}{2\pi}\bigg \{\ln\left(1+e^{-\beta\left[E(0,p_z)-\mu_f\right]}\right)+\ln\left(1+e^{-\beta\left[E(0,p_z)+\mu_f\right]}\right)\bigg \} \, .
 \end{split} \label{Pfree}
\end{align}
\end{widetext}

The one-loop contribution from the gluons has the usual Stefan-Boltzmann form \cite{Kapusta:2006pm}
\begin{equation}
P_{\rm free}^G=2(N_c^2-1)\frac{\pi^2 T^4}{90} \,.
\end{equation}

The two-loop (exchange) contribution from the quark sector can be written as 
\begin{align}
\frac{P_{\rm exch}^{\rm LLL}}{N_c}=-\frac{1}{2}\left(\frac{q_fB}{2\pi}\right)\int \frac{dm_k}{(2\pi)^2}m_k e^{-\frac{m_k^2}{2 q_f B}}
  \mathcal{G}(m_k^2,m_f^2) \, ,
  \label{P_exch_ini}
 \end{align}
where
\begin{widetext}
    \begin{align}
\mathcal{G}(m_k^2,m_f^2)=& g^2 \left(\frac{N_c^2-1}{2}\right)\int\frac{dk_zdp_zdq_z}{(2\pi)^3}(2\pi)\delta(p_z-q_z-k_z)T^3\sum_{\ell,n_1,n_2}\beta\delta_{n_1,n_2+\ell}\frac{4m_f^2}{[\boldsymbol{k}_L^2-m_k^2][\boldsymbol{p}_L^2-m_f^2][\boldsymbol{q}_L^2-m_f^2]},\label{G_ini}  
\end{align}
\end{widetext}
Here $\boldsymbol{k}_L=(i\omega_\ell^B,k_z)$,
$\boldsymbol{p}_L=(i\omega_{n_1}^F+\mu_f,p_z)$,
$\boldsymbol{q}_L=(i\omega_{n_2}^F+\mu_f,q_z)$, where
$\omega_\ell^B=2\pi\ell T$ for bosons and $\omega_{n_i}^F=(2 n_i+1)\pi T$ for
fermions are Matsubara frequencies, with $\ell$ and $n_i$ integers.

Evaluating the momentum integrations yields \footnote{ In this
expression, the UV limit is controlled by the Gaussian weight
$e^{-\frac{m_k^2}{2 q_f B}}$, while the IR sector is controlled by the $m^2_f$
overall factor. As a result, the two-loop pressure in the presence of an
extremely strong magnetic background vanishes in the chiral limit
\cite{Blaizot:2012sd,Fraga:2023lzn}. }
\begin{align}
\begin{split}
\mathcal{G}(m_k^2,m_f^2)=-\beta V g^2\left(\frac{N_c^2-1}{2}\right)T^2 m_f^2
\sum_{\ell,n_2}\frac{\mathcal{E}_\ell-\mathcal{E}_{n_2}}{\mathcal{E}_\ell \mathcal{E}_{n_1} \mathcal{E}_{n_2} \left|\mathcal{E}_\ell-\mathcal{E}_{n_2}\right|\left(\left|\mathcal{E}_\ell-\mathcal{E}_{n_2}\right|+\mathcal{E}_{n_1}\right)} \, ,
\end{split}\label{G_final}
\end{align}
where 

\begin{equation}
\mathcal{E}_\ell=\sqrt{\omega_\ell^2+m_k^2} \,,
\end{equation}
\begin{equation}
\mathcal{E}_{n_1}=\sqrt{(\omega_{n_2}+\omega_\ell+i\mu_f)^2+m_f^2} \,,
\end{equation}
\begin{equation}
\mathcal{E}_{n_2}=\sqrt{(\omega_{n_2}+i\mu_f)^2+m_f^2} \,.   
\end{equation}
As remarked long ago \cite{Blaizot:2012sd}, the chiral limit for
perturbative
QCD at very high magnetic fields is extremely simple: the
exchange contribution vanishes identically for $m_f = 0$ (see also Ref.
\cite{Fraga:2023cef}).

The two-loop contribution from the gluons is given by the well-known formula \cite{Kapusta:2006pm}:
\begin{equation}
P_{\rm 2}^G=-N_c (N_c^2-1)\frac{g^2 T^4}{144} \,,
\end{equation}
so that the total pressure to two-loop order can be written as:
\begin{equation}
P_{\rm 2L}= P_{\rm free}^G + P_{\rm 2}^G + P_{\rm free}^{\rm LLL} +  P_{\rm exch}^{\rm LLL}
\,.
\label{PNLO}
\end{equation}

This expression can be used to compute the coefficients of the Taylor expansion for the pressure in powers of $\hat\mu_B\equiv \mu_B /T$ in the presence of a large magnetic field, such that
\begin{align}
    \frac{P}{T^4}=c_0(T,B)+c_2(T,B)\hat{\mu}_B^2+c_4(T,B)\hat{\mu}_B^4+\mathcal{O}(\hat{\mu}_B^6),
    \label{Taylor}
\end{align}
where the coefficients are
\begin{align}
c_n (T,B) =\frac{1}{n!} \frac{\partial^n P}{\partial \mu_B^n}\bigg{|}_{\mu_B=0}.
\end{align}
Those can be directly compared to recent lattice results \cite{Astrakhantsev:2024mat}, as we show in the sequel.

The pressure to two loops for $3$ flavors with physical quark masses depends not only on the temperature, baryon chemical potential and magnetic field, but also on the renormalization subtraction point $\bar{\Lambda}$, an additional mass scale introduced by the perturbative expansion. This comes about via the scale dependence of both the strong coupling $\alpha_s(\bar{\Lambda})$ and strange quark masses $m_s(\bar{\Lambda})$. 

The running of both $\alpha_s$ and $m_s$ are known to four-loop order in the $\overline{\rm MS}$ scheme \cite{Vermaseren:1997fq}. For our $O(\alpha_s)$ calculation of the QCD pressure, we adopt for the coupling \cite{Fraga:2004gz}
     \begin{equation}
     \alpha_{s}(\bar{\Lambda})=\frac{4\pi}{\beta_{0}L}\left(
     1-\frac{2\beta_{1}}{\beta^{2}_{0}}\frac{\ln{L}}{L}\right) \,,
     \label{eq:alphas}
     \end{equation}
where $\beta_{0}=11-2N_{f}/3$, $\beta_{1}=51-19N_{f}/3$, $L=2\ln\left(\bar{\Lambda}/\Lambda_{\rm \overline{MS}}\right)$. It is clear therefore that the running of $\alpha_{s}$ is a function of $N_{f}$, so that fixing the quark mass at some energy scale also depends on the number of flavors.
For the strange quark mass, we have
	\begin{eqnarray}
	\begin{aligned}
	m_{s}(\bar{\Lambda})=\hat{m}_{s}\left(\frac{\alpha_{s}}{\pi}\right)^{4/9}
	\left[1+0.895062\left(\frac{\alpha_{s}}{\pi}\right)  
 \right] \;,
	\end{aligned}
	\label{eq:smass}
	\end{eqnarray}
with $\hat{m}_{s}$ being the renormalization group invariant strange quark mass, i.e. $\bar{\Lambda}$ independent. Since Eq. (\ref{eq:alphas}) for $\alpha_{s}$ tells us that different values of $N_{f}$ give different values of $\Lambda_{\overline{\rm MS}}$, by choosing  $\alpha_{s}(\bar{\Lambda}=1.5~{\rm GeV},~N_{f}=3)=0.336^{+0.012}_{-0.008}$ \cite{Bazavov:2014soa}, we obtain $\Lambda^{2+1}_{\overline{\rm MS}}=343^{+18}_{-12}~$MeV. Finally, fixing the strange quark mass at $m_{s}(2~{\rm GeV}, N_{f}=3)=92.4(1.5)~$MeV \cite{Chakraborty:2014aca} gives $\hat{m}^{2+1}_{s}~{\approx}~248.7~$MeV when using $\alpha^{2+1}_{s}$ in Eq. (\ref{eq:smass}).

 Following the detailed analysis of different implementations of the scale dependence performed in Ref. \cite{Fraga:2023cef}, we show results only for the most physical case: $\bar{\Lambda}=\sqrt{(2\pi T)^2+eB}$, which corresponds to a natural extension of what is done in finite-temperature field theory in the absence of a magnetic field \cite{Kapusta:2006pm}. Hence, we show results from pQCD including bands that encode the renormalization-scale dependence in the standard range between half the central scale $\bar{\Lambda}$ and twice its value.

\section{Results}
\label{results}

In what follows we present our perturbative results for the first two non-vanishing coefficients of the Taylor expansion for the pressure in powers of $\hat\mu_B\equiv \mu_B /T$. One should keep in mind that, since we adopt the lowest-Landau level approximation in order to obtain analytic results and more control on qualitative aspects, the region of validity for our framework is restricted to $m_s \ll T \ll \sqrt{eB}$, where $m_s$ is the strange quark mass, $e$ is the fundamental electric charge, $T$ is the temperature, and $B$ is the magnetic field strength. So, we show perturbative bands that, presumably, lattice results should reach at higher temperatures, as it does for $c_2$ and $c_4$. 

\begin{figure*}[!ht]
 \centering
 \includegraphics[width=0.65\textwidth]{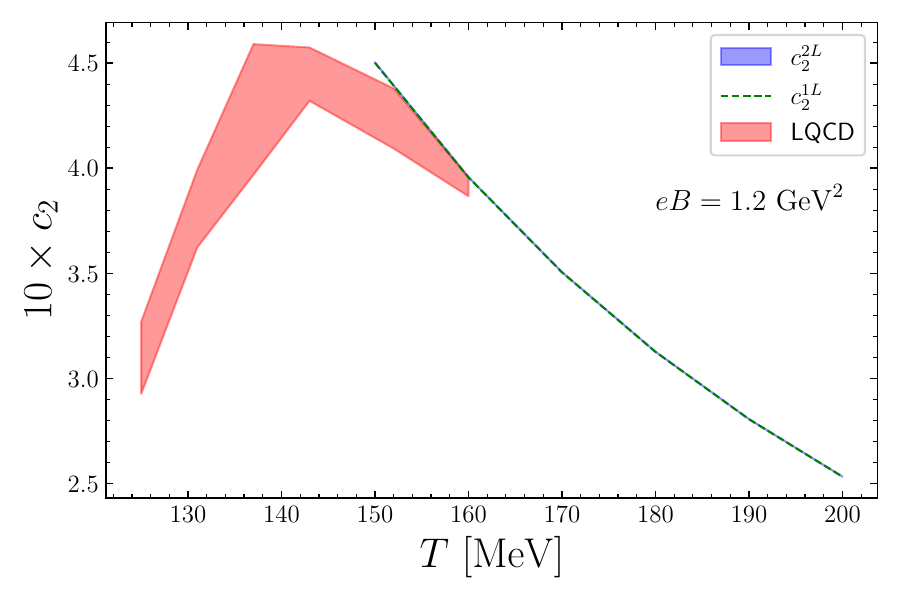} 
 \centering
\caption{Coefficient $c_2$ as a function of the temperature for $eB=1.2$ GeV$^2$. Lattice data (lower, red band) from Ref. \cite{Astrakhantsev:2024mat} and bands on the right from pQCD up to one and two loops. 
Bands in the perturbative part represent the renormalization-scale dependence in the standard range between half the central scale $\bar{\Lambda}$ and twice its value.}
\label{c2_B1_2}
\end{figure*}

In Figure \ref{c2_B1_2} the coefficient $c_2$ as a function of the temperature is displayed for $eB=1.2$ GeV$^2$, comparing our pQCD results with the lattice data from Ref. \cite{Astrakhantsev:2024mat}. 
In this reference, lattice simulations are performed for chemical potentials such that $\mu_u=\mu_d$ and $\mu_s=0$. In our perturbative calculations that corresponds to having no contribution from the $s$ quark to the coefficients $c_n$.  That is precisely the reason why the one-loop contribution, $c_2^{\rm{1L}}$, has no scale dependence, whereas the full two-loop coefficient, $c_2^{2L}$, displays a tiny theoretical uncertainty band: for $u$ and $d$ quarks, only the two-loop contribution has scale dependence and it is very small for small masses, since $c_n\sim m^2$, as can be seen in Eq. (\ref{G_final}). Figure  \ref{c2_B1_2} clearly shows that
the perturbative results seem to be compatible with those from lattice QCD for higher temperatures.

\begin{figure*}[!ht]
 \centering
 \includegraphics[width=0.65\textwidth]{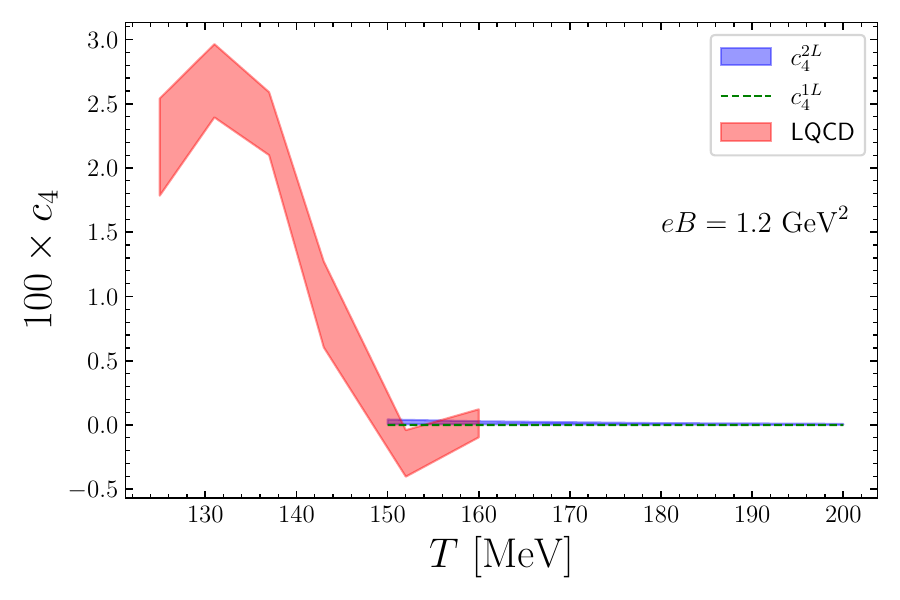} 

 \centering
 
\caption{Coefficient $c_4$ as a function of the temperature for $eB=1.2$ GeV$^2$. Lattice data (left band) from Ref. \cite{Astrakhantsev:2024mat} and bands on the right from pQCD up to one and two loops. 
Bands in the perturbative part represent the renormalization-scale dependence in the standard range between half the central scale $\bar{\Lambda}$ and twice its value.}
\label{c4_B1_2}
\end{figure*}

\begin{figure*}[!ht]

 \centering
 \includegraphics[width=0.65\textwidth]{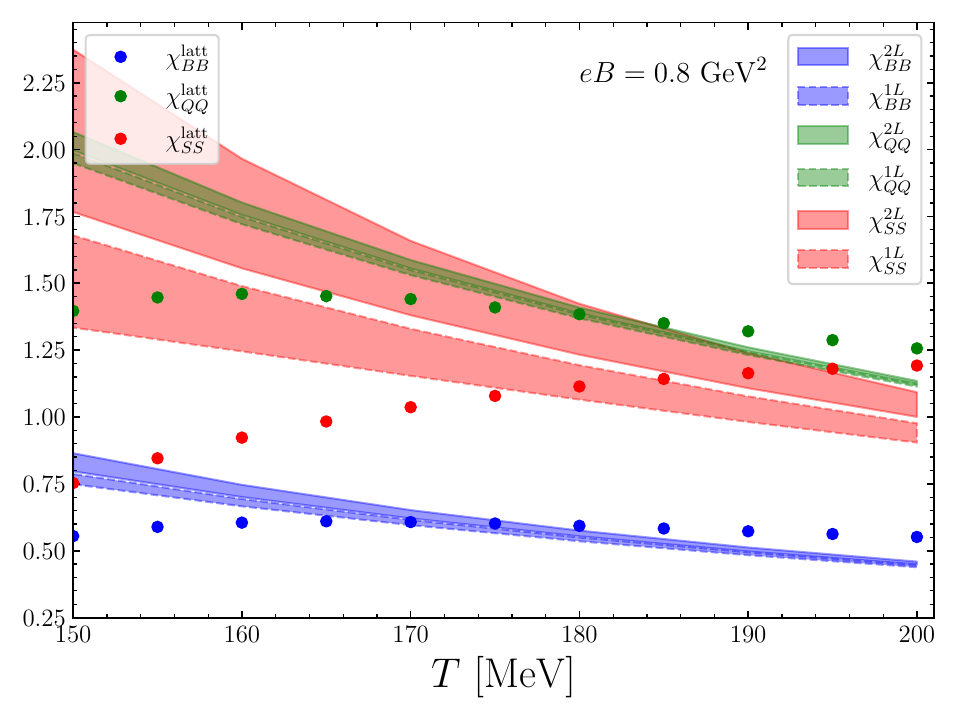} 

\caption{$\chi_{BB}$, $\chi_{QQ}$ and $\chi_{SS}$ as functions of the temperature for  $eB=0.8$ $\rm{GeV}^2$. Lattice data from Ref. \cite{Borsanyi:2023buy}. Bands in the perturbative results represent the renormalization-scale dependence in the standard range between half the central scale $\bar{\Lambda}$ and twice its value. }
\label{chi_B0_8}
\end{figure*}

Figure \ref{c4_B1_2} displays our results for the coefficient $c_4$ as a function of the temperature for $eB=1.2$ GeV$^2$, as well as Lattice data from Ref. \cite{Astrakhantsev:2024mat}. One can see that the lattice band merges quite smoothly onto the (narrow) perturbative band, so that there is a nice overlap region. The perturbative band at two-loop order is narrow, and the dislocation between  one- and two-loop contributions is considerably larger\footnote{This is highly dependent on the mass value. When there are contributions from the strange quark, the exchange contribution is larger, as can be seen in Fig. \ref{chi_B0_8}.} than in the case of $c_2$. However, it still requires magnification of the plot to be visible and remains much smaller than the typical size of the lattice QCD band.

\begin{figure*}[!ht]

 \centering
 \includegraphics[width=0.65\textwidth]{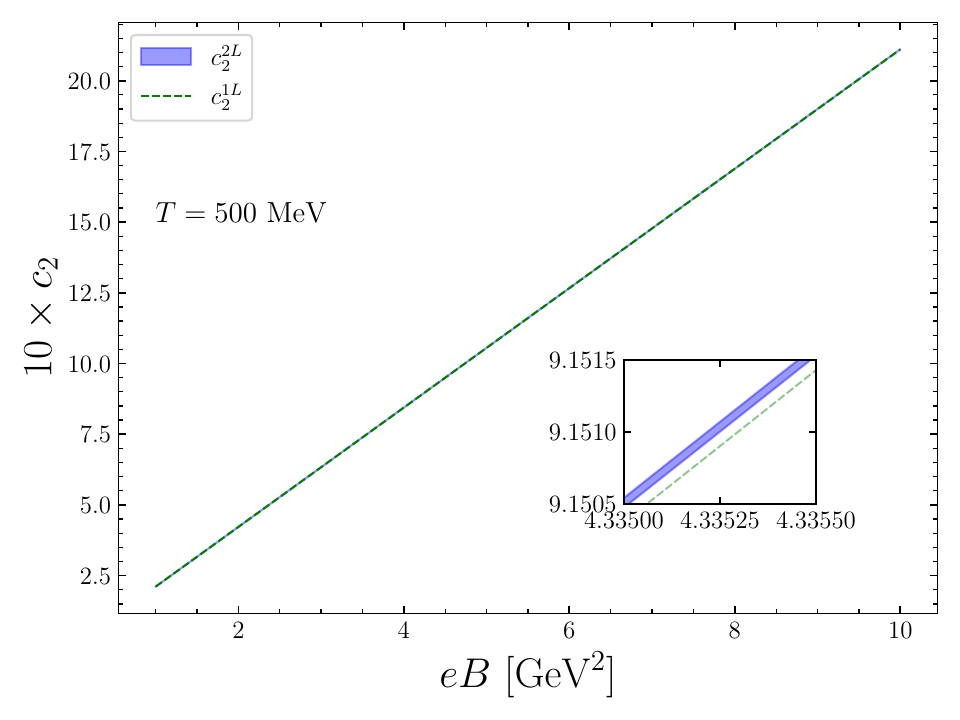} 

 \centering
 
\caption{Coefficient $c_2$ as a function of the magnetic field for $T=500$ MeV. Bands from pQCD up to one and two loops,  
which represent the renormalization-scale dependence in the standard range between half the central scale $\bar{\Lambda}$ and twice its value.}
\label{c2_T5}
\end{figure*}


Our predictions for the behavior of $c_2$ as function of the magnetic field for $T=500$ MeV are shown in Figure \ref{c2_T5}, going to much higher values of the magnetic field strength, up to $10 $ GeV$^2$. As the magnetic field increases, $c_2$ increases in a linear fashion with a very narrow perturbative band. The linear trend is in line with lattice data shown in Ref. \cite{Astrakhantsev:2024mat} for three values of the magnetic field, $eB=1.2$ GeV$^2$ being the highest.
Our perturbative prediction then states that the linear behavior should extend also to much higher magnetic field values. The coefficient $c_4$ for large magnetic fields is, on the other hand, very small, essentially zero.

\begin{figure*}[!ht]
 \centering
 \includegraphics[width=0.65\textwidth]{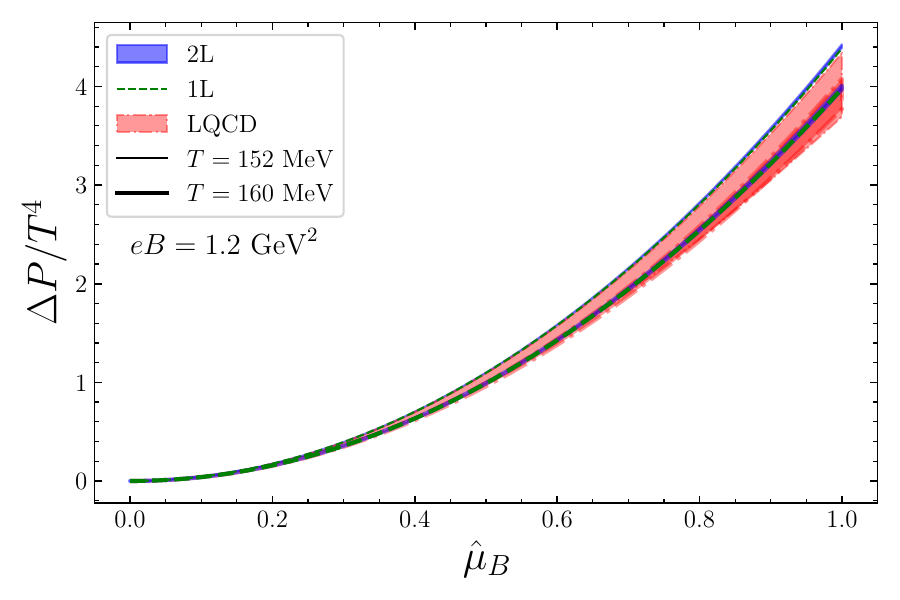} 
 \centering
\caption{The excess of pressure, $\Delta P$, as function of $\hat{\mu}_B$ for two values of temperature, $T= 152, 160$ MeV. Here, $eB=1.2$ $\rm{GeV}^2$. Lattice data from Ref. \cite{Astrakhantsev:2024mat}. Bands in the perturbative results represent the renormalization-scale dependence in the standard range between half the central scale $\bar{\Lambda}$ and twice its value.}
\label{DeltaP}
\end{figure*}

The fact that the coefficient $c_4$ presents a larger contribution from two loops might rise concerns with respect to the convergence of perturbation theory in this setting. However, this is misleading. First of all, it is important to notice that in all cases the coefficient $c_4$ is always at least an order of magnitude smaller than $c_2$. The convergence happens in the expansion in the strong coupling, $\alpha_s$, and in the expansion in (small) $\hat\mu_B$, the former being further reassured by smaller values of the products $c_n (T,B) \hat\mu_B^n$ for higher loop contributions. This can be directly seen in Figure \ref{DeltaP}, where we plot the excess in pressure $\Delta P = P(T,\mu_B)-P(T,\mu_B=0)$ and show that it behaves as expected, with the two-loop result being a small correction with respect to the one-loop contribution. Moreover, the perturbative results are in very good agreement with the lattice predictions.

\begin{figure*}[!ht]

 \centering
 \includegraphics[width=0.65\textwidth]{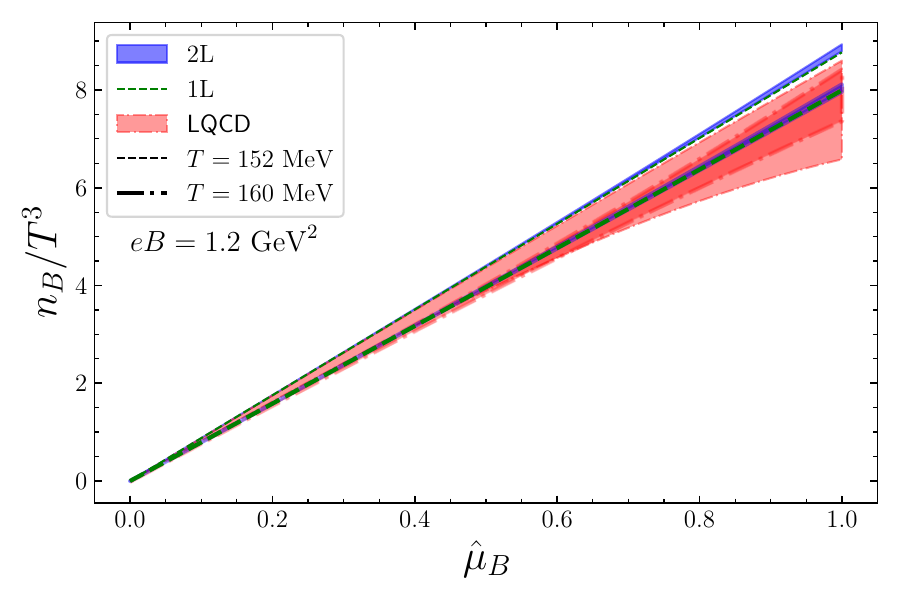} 

\caption{Baryon density as a function of $\hat{\mu}_B$ for two values of temperature, $T= 152, 160$ MeV. Here,  $eB=1.2$ $\rm{GeV}^2$. Lattice data from Ref. \cite{Astrakhantsev:2024mat}. Bands in the perturbative results represent the renormalization-scale dependence in the standard range between half the central scale $\bar{\Lambda}$ and twice its value. }
\label{n_B1}
\end{figure*}


\begin{figure*}[!ht]
 \centering

\includegraphics[width=0.65\textwidth]{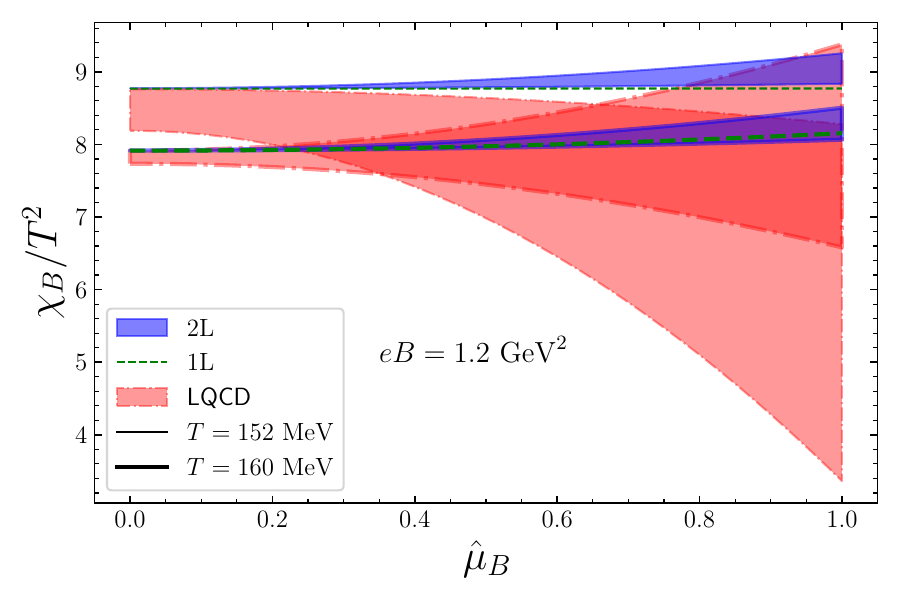} 

\caption{Baryon number susceptibility as a function of $\hat{\mu}_B$ for two values of temperature, $T= 152, 160$ MeV. Here,  $eB=1.2$ $\rm{GeV}^2$. Lattice data from Ref. \cite{Astrakhantsev:2024mat}. Bands in the perturbative results represent the renormalization-scale dependence in the standard range between half the central scale $\bar{\Lambda}$ and twice its value. }
\label{chi_B1}
\end{figure*}


In Figures \ref{n_B1} and \ref{chi_B1} we show the baryon number density and the baryon number susceptibility, respectively, as functions of $\hat{\mu}_B$ for $T=152,160$ MeV and $eB=1.2$ $\rm{GeV}^2$. We include only the contributions from $c_2$ and $c_4$ for both the pQCD expansion and Lattice QCD. 
The perturbative band is very close to that from the Lattice, as expected from the results of Fig \ref{DeltaP}, yet exhibiting a more distinct deviation as $\hat{\mu}_B\to 1$, where the expansion is less accurate.

\newpage

Finally, we present results for the susceptibilities in the $ \left\lbrace \mu_B,\mu_Q,\mu_S \right\rbrace$ basis, which can be compared to the findings of Ref. \cite{Borsanyi:2023buy} on the lattice for a smaller value of the magnetic field. There they are given by the following relations:
 \begin{align}
     \chi_{BB}&=\frac{1}{9}\left(\chi_{200}+\chi_{020}+\chi_{002}+2\chi_{110}+2\chi_{101}+2\chi_{011}\right) \,,\\
     \chi_{QQ}&=\frac{1}{9}\left(4\chi_{200}+\chi_{020}+\chi_{002}-4\chi_{110}-4\chi_{101}+2\chi_{011}\right) \,,\\
     \chi_{SS}&=\chi_{002} \,,
 \end{align}
 where $\chi_{ijk}$ is given by
 \begin{align}
     \chi_{ijk}=\left(\frac{\partial}{\partial \hat{\mu}_u}\right)^i\left(\frac{\partial}{\partial \hat{\mu}_d}\right)^j\left(\frac{\partial}{\partial \hat{\mu}_s}\right)^k \frac{P}{T^4}.
 \end{align}

Our perturbative two-loop calculation is diagonal in flavor, since non-diagonal contributions appear only at the next perturbative order due to the ring resummation \cite{Freedman:1976ub}. Thus, we have $\chi_{110}=\chi_{101}=\chi_{011}=0$.
 In Figure \ref{chi_B0_8} we show the susceptibilities as functions of the temperature for a magnetic field $eB=0.8$ $\rm{GeV}^2$. 
 We compare our perturbative results with lattice data from reference \cite{Borsanyi:2023buy}. Here, we can notice that the strange contribution has a larger scale dependence due to the contribution of $m_s$, as discussed before.
 Even though the magnetic field is probably too low for our LLL hierarchy of scales to hold, it is encouraging that our results are in the same ballpark of Lattice QCD for $T\gtrsim170$ MeV.

It is clear that lattice simulations are not within the ideal
hierarchy of scales to provide a fair comparison to perturbative QCD yet. In
this regard, it would be necessary to have more values of the magnetic field
from the lattice to compare in the future. Nevertheless, our results are in a
reasonably good agreement, which is encouraging. They, of course, miss the
non-monotonic behavior of the coefficients $c_2$ and $c_4$, as expected from a
perturbative calculation. On the other hand, the reduction of the contribution
from higher orders in perturbation theory and the narrow uncertainty bands from
the renormalization scale freedom make the description reasonable.


\section{Summary and outlook}
\label{summary}

In this paper, we computed
the coefficients $c_2(T,B)$ and $c_4(T,B)$ of the Taylor expansion for the pressure in powers of $\mu_B/T$ within perturbative QCD at finite temperature and density, and in the presence of very high magnetic fields, up to two loops with physical quark masses. We also calculated the excess of pressure, baryon density and baryon number susceptibility as functions of $\hat\mu_B$, as well as susceptibilities as functions of the temperature in the $\{ \mu_B,\mu_Q,\mu_S \}$ basis. Since we adopt the lowest-Landau level approximation in order to obtain analytic results and more control on qualitative aspects, the region of validity for our framework is restricted to $m_s \ll T \ll \sqrt{eB}$, where $m_s$ is the strange quark mass, $e$ is the fundamental electric charge, $T$ is the temperature, and $B$ is the magnetic field strength. Even though current lattice results
do not overlap with its region of validity, we found that perturbative results are compatible with those obtained on the lattice for the largest temperatures probed \cite{Astrakhantsev:2024mat}. 

The magnetic perturbative series exhibits mostly narrow error bands, which originate in the truncation of the perturbative series and the mass scale that emerges as a consequence, giving more robustness to the overlap in $c_4$ and the good tendency in $c_2$ when compared to lattice data in their temperature dependence. As for their magnetic field dependence, we provide predictions to $c_2(T,B)$ to larger fields at high temperature, $c_4(T,B)$ being vanishingly small. Those are qualitatively in line with results for three smaller values of the field shown in Ref. \cite{Astrakhantsev:2024mat}. Our findings for the excess of pressure, baryon density and baryon number susceptibility are also in good agreement with lattice data for small values of $\hat\mu_B$. Even for the smaller intensity of magnetic field considered in the lattice simulations of Ref. \cite{Borsanyi:2023buy}, our results for the susceptibilities $\chi_{BB}$, $\chi_{QQ}$ and $\chi_{SS}$ are reasonable when compared to their high end in temperatures. 

The overall results reinforce the conjecture that the perturbative series even in a thermal setting displays a significantly improved behavior in the presence of an extreme magnetic background, at least in the (dominant) quark sector. This is ultimately related to a relative suppression of quark interactions in this regime. A systematic study of the convergence of the perturbative series is certainly required to verify this improvement.
In order to reach conditions closer to phenomenological settings for the physical scales involved, relaxing the lowest-Landau level approximation is probably the most important refinement, representing a feasible albeit technical challenge.

To consider moderate magnetic fields, in a region that is more
relevant phenomenologically, one should probably combine perturbative QCD with a
model description for lower temperatures and densities. Although this is
certainly useful, and a path we plan to follow, one should have in mind that
much lower magnetic fields barely affect the equation of state and the phase
diagram. Moreover, that would depart from the clean first-principle calculation
we have presented in this paper.

\begin{acknowledgments}
E.S.F. thanks A. Yu. Kotov for useful discussions. 
This work was partially supported by CAPES (Finance Code 001), Conselho Nacional de Desenvolvimento Cient\'{\i}fico e Tecnol\'{o}gico (CNPq), Funda\c c\~ao Carlos Chagas Filho de Amparo \` a Pesquisa do Estado do Rio de Janeiro (FAPERJ), and INCT-FNA (Process No. 464898/2014-5). T.E.R acknowledges support from FAPERJ, Process SEI-260003/019683/2022. This work has been supported by STRONG-2020 ``The strong interaction at the frontier of knowledge: fundamental research and applications'' which received funding from the European Union’s Horizon 2020 research and innovation program under grant agreement No 824093.
\end{acknowledgments}
\bibliography{refs.bib}
\end{document}